\documentstyle[eqsecnum,preprint,aps]{revtex}
\begin{document}
\draft
\begin{title} {\Large \bf Quantum fluctuations and glassy behavior:
The case of a quantum particle in a random potential.}
\end{title}
\author{Yadin Y. Goldschmidt}
\address{Department of Physics and Astronomy\\
University of Pittsburgh\\
Pittsburgh, PA  15260}
\date{25 Aug. 1995}
\maketitle
\begin{abstract}
In this paper we expand our previous investigation of a quantum
particle subject to the action of a random potential plus a fixed
harmonic potential at a finite temperature T. In the classical limit
the system reduces to a well-known ``toy'' model for an interface in a
random medium. It also applies to a single quantum particle like an an
electron subject to random interactions, where the harmonic potential
can be tuned to mimic the effect of a finite box.
Using the variational approximation, or alternatively, the limit of
large spatial dimensions, together with the use the replica
method, and are able to solve  the model and obtain its
phase diagram in the $T - (\hbar^2/m)$ plane, where $m$ is the particle's
mass. The phase diagram is similar to that of a quantum spin-glass in
a transverse field, where the variable $\hbar^2/m$ plays the role of
the transverse field. The glassy phase is characterized by
replica-symmetry-breaking. The quantum transition at zero temperature
is also discussed.

\end{abstract}
\pacs{05.30.-d,05.40.+j,75.10.Nr}
\newpage
\section{Introduction}
The effects of quantum fluctuations on phase transitions is a topic of
significant current research. On the other hand transition into a
glassy phase in disordered systems is a topic which is far from
trivial and requires special analytical and numerical techniques. In
particular the behavior of interfaces in disordered systems has been
the subject of many recent papers [1-12]. It is of interest to investigate
the combined effect of disorder and quantum fluctuations that play a
significant role at low temperatures.

In a recent paper \cite{prl}, we have focused on the problem
of a quantum particle in a random potential
with power law correlations plus a fixed harmonic restoring force.
This problem has previously been studied classically
\cite{villain,mp,gb1,engel,yg2}, by using the variational approximation
\cite{mp,gb1,engel}(or alternatively the large $N$-limit
\cite{yg1,yg2}, where $N$ is the number of
dimensions). A transition into a glassy phase has been identified. The
signature of such a transition is the appearance of solutions with
replica-symmetry-breaking (RSB) \cite{binder}. RSB is typically
associated with a
complicated free-energy landscape characterized by many local minima
which are separated by high barriers. A sharp transition for a
single particle must be an artifact of the variational approximation
( or alternatively the large-$N$ limit). In reality this must be a
crossover effect over a range of temperatures.

The ``toy'' model of a particle in a random potential in one dimension
is the simplest example of an interface problem where the location of
the particle stands for the location of an interface between to phases
(like up an down spins) and the random potential stands for quenched
impurities which can pin the interface. Understanding of this problem
is the first step of investigating higher dimensional manifolds. But
the quantum version of the model also apply directly to a particle
like an electron in a dirty metal \cite{lifshits}, where the
strength of the harmonic
potential can be tuned to mimic a finite box which the electron is
confined to, like the situation which occurs in certain mesoscopic
systems.

The question we have had in mind was what were
the effects of quantum fluctuations on this glassy transition. There
are some similarities between this problem and the
quantum spin-glass problem in a transverse field\cite{yglai}. In
that problem, tunneling effects produced by the transverse field
eventually lead, for strong enough field to the destruction of the
spin-glass phase. But this phase when it exists, is still
characterized by replica-symmetry-breaking (RSB), for the infinite
range model.
With this analogy in mind, one should notice though, that the
transition which has been found for a
particle in a random potential \cite{mp,gb1,engel} is of the
Almeida-Thouless type \cite{at} in the sense that it is associated
with the appearance of RSB but not with an order-disorder (spin-glass
like) transition.

In this paper we expand our previous investigation of the quantum
particle, and also correct a typographical error that unfortunately
occurred in
equation (7) of \cite{prl}, propagated into eq. (8) and affected some
of the numerical results. We
find as before that the glassy phase, characterized by RSB persists in
the presence of quantum fluctuations. However, we find that for small
enough particle mass, (m and $\hbar$ enter only in the combination
$\kappa=(m/\hbar^2)$) the glassy phase ceases to exists. Thus the parameter
$1/\kappa$ plays a similar role to the transverse field in the Ising
spin-glass. A schematic phase diagram is depicted in Fig. 1.

\section{Basic definitions}

The density matrix for a particle at finite temperature $T=1/k_B \beta$,
subject to an harmonic potential and a random potential $V$, is given
by the functional integral \cite{feynman}:
\begin{eqnarray}
  \rho({\bf x, x'},U)\ = \ \int_{{\bf x}(0)={\bf x}}^{{\bf x}(U)={\bf
    x'}} [d{\bf x}] \ \exp\left \{-{1 \over \hbar} \int_0^U \left [
{m \dot{{\bf x}}(u)^2 \over 2} \ +\ {\mu {\bf x}(u)^2 \over 2} \ +\
V({\bf x}(u)) \right ] du \right \} ,
\label{densitym}
\end{eqnarray}
where ${\bf x}$ is a $N$-dimensional vector ($N$ is the number of
spatial dimensions), and $U=\beta \hbar$.
The variable $u$ has dimensions of time and is often referred to as the
Trotter dimension. For the diagonal elements of $\rho$ we observe that
the trajectory ${\bf x}(u)$ forms a closed path. In this paper we are
concerned with a random quenched potential $V(x)$ and quantities of
interest are the averaged free energy and the mean-square-displacement
given by
\begin{eqnarray}
  -\beta \langle F\rangle_R \ = \ \left\langle \ln \int \rho ({\bf x}, {\bf
    x}) \ d {\bf x} \right\rangle_R,
 \label{fe} \\
 \langle \langle {\bf x}^2 \rangle \rangle_R \ =\ \left\langle {\int \rho
  ({\bf x}, {\bf x}) \ {\bf x}^2 \ d {\bf
      x} \ \over \ \int \rho ({\bf x}, {\bf x}) \ d {\bf
      x}}\right\rangle_R ,
 \label{x2}
\end{eqnarray}
where, $\langle \rangle_R$ stands for an average over the
random realizations of the potential. The difficulty in carrying out
the quenched averages stems from the fact that in equation (\ref{fe})
the average is taken after taking the logarithm, i.e. the averaged
free energy is not the logarithm of the averaged partition function.
Similarly in eq. (\ref{x2}) the average of the quotient is not equal to
the quotient of the averages.
We take the potential $V(x)$ to gaussian distributed, which means that
the probability for a given realization of the potential is
\begin{eqnarray}
P(V(x)) = C \exp (-\int dx dx' V(x) \Delta (x-x') V(x') ) ,
\label{vgauss}
\end{eqnarray}
with some known function $\Delta(x-x')$.
It is thus sufficient to know only the first two moments of the
distribution vis.
\begin{eqnarray}
\left<V({\bf x})\right>_R = 0, \ \left<V({\bf x})V({\bf
    x'})\right>_R = - N f \left({({\bf x}-{\bf x'})^2 \over N
    }\right).
\label{potav}
\end{eqnarray}
where the functions $f$ and $\Delta$ are related to each other.
The function $f$ describing the correlations of the random potential
is taken to decay as a power at large distances:
\begin{eqnarray}
  f(y)\ = \ {g \over 2(1-\gamma)}\ \ (a_0 \ +\ y)^{\ 1-\gamma}.
\label{f}
\end{eqnarray}
The index $\gamma$ describes the behavior of the correlations of the
disorder at large distances. In this paper we consider only two cases,
that of $\gamma=3/2$ which we call the case of short-ranged
correlations, and that of $\gamma=1/2$ which we call the case of
long-ranged correlations \cite{mp1,mp,engel}. The parameter $a_0$
plays the role of a short-distance regulator for $f$. We chose these
values of $\gamma$ in order to make contact with known results in the
classical case. In the classical case it has been shown (for
$N=1$), that faster falling correlations than those characterized by
$\gamma=3/2$ , even correlations falling exponentially fast, are
equivalent within the variational approximation to the case of $\gamma=3/2$
\cite{mp1,engel}. This fact also holds in the quantum case, as is
demonstrated in sec. III below. The long-range case is of interest
because of its connection with the directed-polymer problem and with
the random-field Ising model \cite{parisi,mp1,gb1,engel}.

We apply the replica trick in order to carry out the quenched
average over the random realizations. We consider $n$-copies of the
system, and obtain for the averaged density matrix:
\begin{eqnarray}
\rho({\bf x}_1 \cdots {\bf x}_n, {\bf x}_1 \cdots {\bf x}_n,U)\ = \
\int_{{\bf x}_a(0)={\bf x}_a}^{{\bf x}_a(U)={\bf x}_a}\ \prod_{a=1}^n
 [d{\bf x}_a] \exp \left \{-{\cal H}_n/\hbar \right \},\\
{\cal H}_n\ = \ {1 \over 2} \int_0^U du\ \sum_a \left[ m
 {\bf \dot{x}}_a^2(u) + \ \mu
{\bf x}_a^2(u) \right] \hspace{1.75in} \nonumber \\
+ \ {1 \over 2\hbar}\int_0^U du \int_0^U du'
\sum_{ab} N f\left({({\bf x}_a(u)-{\bf x}_b(u'))^2 \over N} \right).
\label{Hn}
\end{eqnarray}
Note that this equation although somewhat similar, is quite different
from the equation for the $n$-body hamiltonian corresponding to the
directed polymer problem in $1+N$ dimensions \cite{mp1}. The
difference is first
the fact that the integral over the ``time'' variable $u$ is on a finite
interval, and second and more significantly, that the $n$-body
potential is non-local in ``time'', i.e. it involves a double integral
over both $u$ and $u'$.

\section{The large-$N$ limit and the variational approximation}

In the large-$N$ limit, we introduce collective variable fields
$r_{ab}(u,u')$:
\begin{eqnarray}
r_{ab}(u,u')= {1 \over N} {\bf x}_a(u) \cdot {\bf x}_b(u')
  \label{rab}
\end{eqnarray}
and Lagrange multiplier fields $s_{ab}(u,u')$ to implement (\ref{rab}).
We find \cite{yg1}
\begin{eqnarray}
\rho({\bf x}_1 \cdots {\bf x}_n, {\bf x}_1 \cdots {\bf x}_n,U)\ = \
\int_{{\bf x}_a(0)={\bf x}_a}^{{\bf x}_a(U)={\bf x}_a}\ \prod_{a=1}^n
 [d{\bf x}_a] \int \prod_{a \leq b}
 [dr_{ab}(u,u')][ds_{ab}(u,u')]  \nonumber \\
\exp \left \{- {1 \over 2 \hbar^2} \int_0^U du \int_0^U du'
\sum_{ab}(N r_{ab}(u,u') - {\bf x}_a(u) \cdot {\bf
x}_b(u'))s_{ab}(u,u') \right.  \nonumber \\
-{1 \over 2\hbar} \int_0^U du\ \sum_a \left[ m
 {\bf \dot{x}}_a^2(u)  + \mu  {\bf x}_a^2(u) \right] \nonumber \\
-\left. \ {1 \over 2\hbar^2}\int \int_0^U du du'
\sum_{ab} N f\left(r_{aa}(u,u)+r_{bb}(u',u')-2 r_{ab}(u,u') \right)
\right \}.
  \label{rho&r&s}
\end{eqnarray}
{}From here it follows that the free energy is given by:
\begin{eqnarray}
n \beta \left< F \right>_R /N \ = \ {1 \over 2 \hbar^2} \int_0^U du
\int_0^U du' \sum_{ab}^n \left( r_{ab}(u,u') s_{ab}(u,u') \ + \right.
\hspace{1.5in} \nonumber \\   \left.
f(r_{aa}(u,u)+r_{bb}(u',u')-2r_{ab}(u,u'))\right)-\ln
\int_{-\infty}^\infty \prod_a dx_a \int_{x_a(0)=x_a(U)=x_a} \ \prod_a
[dx_a] e^J,
\label{F&J}
\end{eqnarray}
where we defined
\begin{eqnarray}
J= -{1 \over 2\hbar} \int_0^U du \sum_a \left( m
\dot{x}_a^2(u) + \mu
x_a^2(u)\right) \ + \ {1 \over 2\hbar^2}\int\int_0^U du\ du'\sum_{ab}
s_{ab}(u,u')x_a(u)x_b(u').
\label{J}
\end{eqnarray}
The limit $n \rightarrow 0$ is to be taken.
In equation (\ref{F&J}) the variables $x_a(u)$ are scalers since a
factor of $N$ has been extracted. In the large $N$ limit  the collective
variables $r_{ab}(u,u')$ and $s_{ab}(u,u')$ are determined by
the stationarity of the free-energy. We will be looking for solutions
of the saddle-point equations obeying translational
invariance in the ``time'' direction, i.e. such that the
order-parameters depend only on the difference $\zeta=\ u-u'$. They
should also be periodic functions of this variable with period $U$,
and also symmetric under $\zeta \rightarrow -\zeta$ (even functions).

In the case of a classical particle there is another approach to the
problem called the variational approximation which yields essentially
the same results as the large-$N$ limit aside from a simple
renormalization of the correlation function of the disorder. For the
quantum potential we were able to show that the same results still
hold. Let us introduce the variational hamiltonian:
\begin{eqnarray}
h_n\ = \ {1 \over 2} \int_0^U du\ \sum_a \left[ m
 {\bf \dot{x}}_a^2(u) + \ \mu
{\bf x}_a^2(u) \right] \hspace{1.75in} \nonumber \\
- \ {1 \over 2\hbar}\int_0^U du \int_0^U du'
\sum_{ab} s_{ab}(u-u'){\bf x}_a(u)\cdot {\bf x}_b(u').
  \label{hn}
\end{eqnarray}

Then the variational free-energy is given by:
\begin{eqnarray}
n \beta \left< F \right>_R /N \ = \ \langle {\cal H}_n -h_n\rangle_{h_n}
/\hbar - \ln \int [dx]e^{-h_n/\hbar}.
  \label{fvar}
\end{eqnarray}

In Appendix A we show that the variational free-energy coincides with
that given in the large-$N$ limit (see eq.(\ref{fe2}) below), with the
following renormalization of the function $f $ defined in
eq. (\ref{f}):
\begin{eqnarray}
f(y) \rightarrow \hat{f}(y)={1 \over \Gamma(N/2)}\int_0^\infty d\alpha
\alpha^{N/2-1} e^{-\alpha} f({2 \alpha \over N} y).
  \label{fhat}
\end{eqnarray}
The effective correlation $\hat{f}$ is similar to $f$ in the sense
that for large arguments:
\begin{eqnarray}
\hat{f}(y) \sim {\hat{g} \over 2(1-\hat{\gamma})}y^{1-\hat{\gamma}},
  \label{assym}
\end{eqnarray}
with
\begin{eqnarray}
\hat{\gamma}=\gamma \hspace{0.25 in} {\rm if} \hspace{0.25 in} \gamma
\leq {N \over 2}+1 \nonumber \\
\hat{\gamma}={N \over 2}+1 \hspace{0.25 in} {\rm if} \hspace{0.25 in} \gamma
\geq {N \over 2}+1
  \label{gammahat}
\end{eqnarray}
and
\begin{eqnarray}
\hat{g}=g {\Gamma(1-\gamma+N/2)\over \Gamma(N/2)}\left({N \over
  2}\right)^{-1+\gamma }.
  \label{ghat}
\end{eqnarray}
Since we will be mainly concerned with  the case where
$\hat{\gamma}=\gamma$, the the values for $g$ quoted in the numerical
work below should actually pertain to the renormalized coupling
$\hat{g}$. In the limit $N \rightarrow \infty$, $\hat{f}$ approaches
$f$.

\section{The stationarity conditions and their exact solution assuming
replica symmetry}

In this section we will derive the stationarity equations for the free
energy and solve them in the replica-symmetric (RS) case. The case of
RSB will be discussed in the following section.
We proceed by transforming to frequency space by using a Fourier series
representation of the order parameters:
\begin{eqnarray}
r_{ab}(\zeta)= {1 \over \beta}\sum_{l=-\infty}^\infty \exp(-i \omega_l
\ \zeta)\ \tilde{r}_{ab}(\omega_l)\\
s_{ab}(\zeta)= {1 \over \beta}\sum_{l=-\infty}^\infty \exp(-i \omega_l
\ \zeta)\ \tilde{s}_{ab}(\omega_l),
  \label{FT}
\end{eqnarray}
where
\begin{eqnarray}
\omega_l = {2 \pi \over U} l, \ \ l=0, \pm 1, \pm2, \cdots \ \ .
  \label{omega}
\end{eqnarray}
The last equation follows from the periodicity of the order
parameters. From the fact that $r_{ab}(\zeta)$ (and also $s$) are even
functions it follows that
$\tilde{r}_{ab}(\omega)=\tilde{r}_{ab}(-\omega)$,
and similarly for $s$.
To proceed we note that for any periodic function  $g(u-u')$ with period
$U$ it follows that
\begin{eqnarray}
\int_0^U du \int_0^U du' g(u-u') = \int_0^U du \int_0^u d\zeta
g(\zeta) + \int_0^U du \int_{-U+u}^0 d\zeta g(\zeta)= \nonumber \\
\int_0^U du \int_0^u d\zeta
g(\zeta) + \int_0^U du \int_{u}^U d\zeta g(\zeta-U)=\int_0^U du
\int_0^U d\zeta g(\zeta) = U \int_0^U d\zeta g(\zeta).
\label{periodic}
\end{eqnarray}
In terms of the new variables introduced in eq. (\ref{FT}), the free energy
becomes:
\begin{eqnarray}
n \beta \left< F \right>_R /N \ = \ {1 \over 2} \sum_\omega \sum_{ab}
\tilde{r}_{ab}(\omega) \tilde{s}_{ab}(\omega) -{1 \over 2} \sum_\omega
{\rm tr}\ \ln G(\omega) \nonumber \\
 + {\beta \over 2 \hbar} \sum_{ab} \int_0^U d \zeta \
 f \left({1 \over \beta} \sum_{\omega'}\
(\tilde{r}_{aa}(\omega')+ \tilde{r}_{bb}(\omega')-
2 e^{-i\omega'\zeta} \ \tilde{r}_{ab}(\omega'))\right) + const.,
\label{fe2}
\end{eqnarray}
where we defined
\begin{eqnarray}
G_{ab}(\omega) \equiv ([(m \omega^2+\mu){\bf 1}-
\tilde{s}(\omega)]^{-1})_{ab}.
\label{G}
\end{eqnarray}
The constant in eq. (\ref{fe2}) can most easily be determined from the
known free energy for a free particle ($f \equiv 0$ and $\mu=0$).

{}From eq. (\ref{fe2}) it follows that stationarity equations are given by:
\begin{eqnarray}
\tilde{r}_{ab}(\omega) = G_{ab}(\omega), \hspace{2in} \label{r-sad}\\
\tilde{s}_{ab}(\omega) = {2 \over \hbar} \int_0^U d\zeta \
e^{i \omega \zeta} f' \left({1 \over \beta} \sum_{\omega'}\
(\tilde{r}_{aa}(\omega')+ \tilde{r}_{bb}(\omega')-
2 e^{-i\omega'\zeta}\tilde{r}_{ab}(\omega'))\right) \hspace{.25in} a
\neq b, \label{s-off}\\
\tilde{s}_{aa}(\omega)+\sum_{b \neq a}\tilde{s}_{ab}(0)
+ {2 \over \hbar} \int_0^U d\zeta \ (1-e^{i \omega\zeta}) f'({2 \over
  \beta} \sum_{\omega'} \tilde{r}_{aa}(\omega')\ (1-e^{-i\omega'\zeta})) = 0.
  \label{s-on}
\end{eqnarray}
In eq. (\ref{s-off}), $f'$ stands for the first derivative of the
function $f$, characterizing the correlations of the random potential.

We proceed to look for a RS solution to these equations by taking
all the elements $\tilde{r}_{ab}(\omega) \equiv \tilde{r}(\omega)$ with
$a \neq b$ to be equal to each
other (for the same frequency) and similarly for
$\tilde{s}_{ab}(\omega) \equiv \tilde{s}(\omega)$.
The diagonal elements are also taken equal to each other, and are
denoted by $\tilde{r}_d(\omega)$ and $\tilde{s}_d(\omega)$.
In the limit $n \rightarrow 0$ we find from equations
(\ref{s-on}), (\ref{s-off}) by inverting an $n \times n$ matrix:
\begin{eqnarray}
\tilde{r}_d(\omega)={1 \over {m \omega^2+\mu-\tilde{s}_d(\omega) +
    \tilde{s}(\omega)}}+{\tilde{s}(\omega) \over (m
\omega^2+\mu-\tilde{s}_d(\omega) + \tilde{s}(\omega))^2}, \\
\tilde{r}(\omega)={\tilde{s}(\omega) \over (m
  \omega^2+\mu-\tilde{s}_d(\omega) + \tilde{s}(\omega))^2}.
\label{inv}
\end{eqnarray}
These expressions have to be substituted into eqs. (\ref{s-off}) and
(\ref{s-on}) to obtain the self-consistent equations for $\tilde{s}$ and
$\tilde{s}_d$:
\begin{eqnarray}
\tilde{s}(\omega) = {2 \over \hbar} \int_0^U d\zeta \
e^{i \omega \zeta} f' \left({2 \over \beta} \sum_{\omega'}\
({1 \over {m \omega'^2+\mu-\tilde{s}_d(\omega') +
    \tilde{s}(\omega')}}+{\tilde{s}(\omega')\ (1-e^{-i\omega'\zeta})
    \over (m  \omega^2+\mu-\tilde{s}_d(\omega) +
    \tilde{s}(\omega))^2})\right),
  \label{s-eq}\\
\tilde{s}_d(\omega)=\tilde{s}(0)
- {2 \over \hbar} \int_0^U d\zeta \ (1-e^{i \omega\zeta})\times \hspace{3.85
  in} \nonumber \\
f'({2 \over \beta} \sum_{\omega'} \ (1-e^{-i\omega'\zeta}) {1 \over {m
    \omega'^2+\mu-\tilde{s}_d(\omega') +
    \tilde{s}(\omega')}}+{\tilde{s}(\omega') \over (m
\omega'^2+\mu-\tilde{s}_d(\omega') + \tilde{s}(\omega'))^2}).
\label{sd-eq}
\end{eqnarray}
We will now seek a solution with time-independent off-diagonal
elements. This is in analogy with quantum spin-glasses
\cite{yglai,sachdev}, the rational
being that the off-diagonal elements of $r$ and $s$ are the glass order
parameters and hence are constants independent of the Trotter time. We
find that such a solution satisfies the saddle point equations
exactly. The diagonal elements are still time-dependent. In frequency
space we look for solutions of the form:
\begin{eqnarray}
\tilde{s}(\omega)=\tilde{s}\ \delta_{\omega, 0}.
  \label{static}
\end{eqnarray}
This ansatz satisfies eqs. (\ref{s-eq}), and we obtain:
\begin{eqnarray}
  \tilde{s}=2\beta f'\left({2 \over \beta \mu}+ {2 \over \beta}
  \sum_{\omega \neq 0} {1 \over m
    \omega'^2+\mu-\tilde{s}_d(\omega')}\right)\\
\tilde{s}_d(\omega)=\tilde{s}
- {2 \over \hbar} \int_0^U d\zeta \ (1-e^{i \omega\zeta})\
f'\left({2 \over \beta} \sum_{\omega' \neq 0} \ (1-e^{-i\omega'\zeta})
{1 \over m \omega'^2+\mu-\tilde{s}_d(\omega') }\right).
\label{s-rs}
\end{eqnarray}

Two physical quantities of interest are
\begin{eqnarray}
\langle \langle{\bf x}^2 \rangle \rangle_R / N = {1 \over n}
\sum_{a=1}^n r_{aa}(0)=
r_d(0)= {1 \over \beta}\sum_{k=-\infty}^\infty \tilde{r}_d (\omega_k),
\label{x2d1}\\
\langle \langle{\bf x}^2 \rangle - \langle{\bf x}\rangle^2 \rangle_R /
N = {1 \over n}\sum_{a=1}^n r_{aa}(0)- {1 \over n(n-1)} \sum_{a \neq b}^n
r_{ab}(0) =r_d(0)-r(0).
  \label{x2d}
\end{eqnarray}
The very last equality holds if one assumes replica symmetry.

\begin{eqnarray}
\langle \langle{\bf x}^2 \rangle \rangle_R / N = {1 \over \beta \mu}+
{1 \over \beta} \sum_{\omega \neq 0} {1 \over m \omega^2 +\mu
  -\tilde{s}_d(\omega)}  + {\tilde{s} \over \beta \mu^2 }
\label{x2f1}
\end{eqnarray}
and
\begin{eqnarray}
\langle \langle{\bf x}^2 \rangle - \langle{\bf x}\rangle^2 \rangle_R /
N = {1 \over \beta \mu} + {1 \over \beta} \sum_{\omega \neq 0}
{1 \over m \omega^2 +\mu -\tilde{s}_d(\omega)}.
  \label{x2f2}
\end{eqnarray}
In all the equations derived above, $m$ and $\hbar$ enter only through
the combination
\begin{equation}
\kappa \equiv {m \over \hbar^2} \ \ .
\label{kappa}
\end{equation}
The classical limit is given by $\hbar \rightarrow 0$ and
consequently $\kappa \rightarrow \infty$. Practically, this limit can
be achieved by taking the particle's mass to be very large. On the
other hand, the quantum limit is obtained for very small particle
mass.
In the classical limit only the zero frequency component of the
observables survives and we obtain e. g.
\begin{eqnarray}
\langle \langle x^2 \rangle\rangle_R = {1 \over \beta \mu}
+{2 \over\mu^2 }f'\left( {2 \over\beta \mu } \right),
\label{x2classical}
\end{eqnarray}
which agrees with refs. \cite{mp,engel}.

{}From equations (\ref{x2d1}) and (\ref{s-rs}) it follows that the
expression for $\langle \langle x^2 \rangle\rangle_R$ in the quantum
case, is the same as in the classical case but with a renormalized
temperature $1/\beta_R$ given by :
\begin{eqnarray}
{1 \over \mu  \beta_R }= {1 \over \mu \beta} + {1 \over \beta}
  \sum_{\omega \neq 0} {1 \over m \omega'^2+\mu-\tilde{s}_d(\omega')}=
{1 \over \mu \beta} + {1 \over 2}b_0(\beta,\kappa,\mu,g),
  \label{betar}
\end{eqnarray}
where we defined:
\begin{eqnarray}
b_0(\beta,\kappa,\mu,g)={2 \over \beta}\sum_{\omega \neq 0} {1 \over m
\omega^2 +\mu  -\tilde{s}_d(\omega)}.
  \label{b0}
\end{eqnarray}
The minimum of the mean-square-displacement in eq.(\ref{x2classical}) is
attained for
\begin{eqnarray}
T_c^{CL}=1/\beta_c^{CL}={1 \over 2} (2\gamma\ g)^{1 \over 1+\gamma}
\mu^{-{1-\gamma \over 1+\gamma}}-{1 \over 2}a_0 \mu
\label{Tccl}
\end{eqnarray}
In the quantum case, it follows from eq.(\ref{betar}), that the
temperature for which the minimum is attained, is given by a solution of the
equation
\begin{eqnarray}
T_c + {1 \over 2}\mu b_0(1/T_c,\kappa,\mu,g)=T_c^{CL}
\label{Tc}
\end{eqnarray}
We will see in the next section that for the case of long-ranged
correlations the temperature for which the minimum of the
mean-square-displacement is attained is indeed the transition
temperature into the glassy state characterized by RSB.
For the case of short ranged correlation of the disorder, even in the
classical case the transition into a replica-broken phase occurs only
approximately at the temperature given by eq.(\ref{Tccl}) if $\mu$ is
not too small \cite{engel}. Otherwise the transition $T^*$ into an RSB
 phase occurs at a higher temperature than given by (\ref{Tccl}).
The value of the mean-square-displacement at the minimum is given by
\begin{eqnarray}
\langle \langle{\bf x}^2 \rangle \rangle_R / N =
(1+\gamma)(2\gamma)^{-{\gamma \over 1+\gamma}}g^{1 \over
1+\gamma}\mu^{-{2 \over 1+\gamma}} - {1 \over 2} a_0.
\label{x2m}
\end{eqnarray}

We now turn to the numerical solution of equations (\ref{s-rs}).
Some technical details of the numerical procedure are given in Appendix B.
First we consider the case of long-ranged correlations of the disorder
with $\gamma=1/2$.
In Fig. 2 we display the results for the mean-square-displacement for
various values of $\kappa=m/\hbar^2$.
We have chosen $g=2 \sqrt{2}$, $a_0=0.01$ and $\mu=1$ in order to
compare with results of the classical case \cite{mp}.
For these values of the parameters, $T_c(\infty)=0.995$.  For large $\kappa$
there is good agreement with known results for the classical particle
\cite{mp}. As $\kappa$ decreases and quantum effects become more
significant, the minimum of the curves moves to the left which shows
that $b_0$ is
increasing and hence $T_c$ is decreasing (see eq. (\ref{Tc}).
For $\kappa \approx 0.17$ the minimum of the curve is at about $T=0$,
and for lower values of $\kappa$ no minimum occurs at a positive temperature.
{}From this data we can reconstruct an approximate phase diagram in the
$T-1/\kappa$ plane, see Fig. 3.

We now turn to the case of short-ranged correlations of the disorder,
i.e. $\gamma=3/2$ (recall eq. (\ref{f})). We have chosen $a_0=1$,
$g=1/\sqrt{2\pi}$ and $\mu=0.53$ in order to compare with the results
of the classical case \cite{engel}. In Fig. 4 we plot the
mean-square-displacement given by eq. (\ref{x2f1}) for various values
of $\kappa$ as a function of $t=T/T_c(\infty)$,
where $T_c(\kappa)$ is the temperature for which the
``transition'' into the glassy state occurs. Thus t=1 is the rescaled
temperature for which the transition occurs in the classical case
($\kappa=\infty$). For the values of the parameters we have chosen,
$T_c(\infty)=0.2082$.
We see that the situation is similar to the long-ranged case. The
value of $\kappa$ for which the minimum of the curve hits zero is
about $2$. The reconstructed phase diagram is given in Fig. 5.

In Appendix III we review, correct and present more details of the
real-space approach used in
our previous investigation \cite{prl}. The formalism is set up for
first-step RSB. (The need for RSB at low temperature will be discussed
in the next section). We have checked numerically that for replica symmetry and
long-ranged correlations of the disorder the plots of the mean-square
displacement vs. temperature coincide with those obtained from the
momentum space calculations as reported above.

We finish this section with a discussion of the zero temperature
limit. In this limit the frequency becomes a continuous variable, and
the equations for $\tilde{s}$ and $\tilde{s}_d(\omega)$ become:
\begin{eqnarray}
\tilde{s}/\beta= 2 f'({1 \over \pi} \int_{-\infty}^\infty {d\omega'
  \over \kappa
\omega'^2+\mu -\tilde{s}_d(\omega'/\hbar)}),\hspace{2.8
in}\label{st0}\nonumber\\
\tilde{s}_d(\omega/\hbar)= -2 \int_{-\infty}^\infty d \zeta (1 -
e^{i\omega\zeta}) \times \hspace{3 in}\nonumber\\
\left[ f'({1 \over \pi}\int_{-\infty}^\infty d\omega' {1 -
  e^{-i\omega'\zeta} \over \kappa
\omega'^2+\mu -\tilde{s}_d(\omega'/\hbar)})- f'({1 \over \pi}
\int_{-\infty} ^\infty
{d\omega' \over \kappa \omega'^2+\mu -\tilde{s}_d(\omega'/\hbar)})\right],
\hspace{0.2 in}\omega \neq 0 \nonumber \\
\tilde{s}_d(0)=\tilde{s}.\hspace{4.5 in}
\label{sdt0}
\end{eqnarray}
where we have rescaled the variables $\zeta$ and $\omega$ by $1/\hbar$
and $\hbar$ respectively. In terms of the solution to the last
equation, the expression for the mean square displacement become:
\begin{eqnarray}
\langle \langle{\bf x}^2 \rangle \rangle_R / N = {1 \over
  2}b_0(\beta=\infty,\kappa,\mu,g) +2 f'(b_0(\beta=\infty,\kappa,\mu,g)).
\label{x2t0}
\end{eqnarray}
with
\begin{eqnarray}
b_0(\beta=\infty,\kappa,\mu,g)={1 \over \pi}\int_{-\infty}^\infty
{d\omega' \over \kappa  \omega'^2+\mu -\tilde{s}_d(\omega'/\hbar)}
  \label{b0inf}
\end{eqnarray}
Eq. (\ref{x2t0}) is similar to the classical expression (\ref{x2f1}),
with the temperature variable $T$ replaced by $(1/2) \mu
b_0(\beta=\infty,\kappa,\mu,g)$.
{}From eq. (\ref{Tc}) it follows that the quantum transition at $T=0$
(for the case of long-ranged correlations) occurs when $\kappa=\kappa_c$,
where
\begin{eqnarray}
{1 \over 2}b_0(\beta=\infty,\kappa_c,\mu,g)= {T_c^{CL} \over \mu}.
\label{kapc}
\end{eqnarray}

\section{Replica-Symmetry-Breaking solution.}

In this section we proceed to explore the possibility of RSB. The need
to break replica symmetry has been demonstrated in the classical case
below a certain transition temperature $T_c$. The physical origin of
RSB is the existence of many local minima of the free energy,
separated by barriers. A sharp transition temperature for a single
particle exists only in the large-$N$ limit or within the framework of
the variational (gaussian) approximation. Classically the pattern of
RSB depends on the range of correlations of the random potential. for
short range correlations a one-step RSB has been found sufficient,
whereas for long-ranged correlations a continuous RSB has been found
necessary \cite{mp,engel}. In the quantum case we will show below,
that RSB also occurs if the strength of the quantum fluctuations is not too
large. Support for this assertion also comes from the fact that
for the RS solution a plot of $\langle \langle x^2 \rangle \rangle_R$
vs. temperature (Fig. 2) is not monotonic, but increases  as
$T \rightarrow 0$ for $\kappa$ not too small. This indicates that the
RS solution is inadequate at low temperatures in this
case. Practically one can compare the free energies associated
with the RS and RSB solutions (if the latter exists) and verify that
the free energy of the RSB solution is {\bf higher} ( an artifact of
the $n \rightarrow 0$ limit). This indicates that RS has to be broken.
This is indeed the situation in the classical case and it carries over
to the quantum case as explained below.

In constructing a RSB solution we associate
\begin{eqnarray}
\tilde{s}_{aa}(\omega)=\tilde{s}_d(\omega), \\
\tilde{s}_{ab}(\omega,z) \leftrightarrow \tilde{s}(z) \ \delta_{\omega,
  0}\ \ ,
  \label{matrix}
\end{eqnarray}
where the Parisi parameter $0<z<1$ labels the 'distance' between the replica
indices $ab$. Again, we have used the static ansatz for the
non-diagonal matrix elements, which will turn out to be a consistent
solution. One should not confuse the frequency dependence of the
diagonal element $\tilde{s}_d(\omega)$ with the z-parameter dependence
of $\tilde{s}(z)$. A similar parametrization applies to
$\tilde{r}_{ab}(\omega)$.
To write down the expressions for $\tilde{r}$, analogous to eqs. (\ref{inv}),
we need the expression
for the inverse of a hierarchical matrix (see Appendix II of \cite{mp}).
We then find:
\begin{eqnarray}
\tilde{r}_d(\omega) = {1 \over m
\omega^2+\mu-\tilde{s}_d(\omega)+\delta_{\omega,0}\int_0^1dz\ \tilde{s}(z)}
\left(1 + {\tilde{s}(z=0)\delta_{\omega,0} \over \mu}\ + \
\delta_{\omega,0} \int_0^1{dz \over z^2}\ {[\tilde{s}](z) \over
\mu+[\tilde{s}](z)}\right) ,
\label{rdz}\\
\tilde{r}(\omega,z)=\tilde{r}(z)\delta_{\omega,0}=\delta_{\omega,0}\ {1
  \over \mu}\left( {[\tilde{s}](z) \over  z(
\mu+[\tilde{s}](z))}\ + \ {\tilde{s}(z=0) \over \mu}\ + \
 \int_0^z{dz \over z^2}\ {[\tilde{s}](z) \over
\mu+[\tilde{s}](z)}\right),
\label{rz}
\end{eqnarray}
where
\begin{eqnarray}
 [\tilde{s}](z) = z \ \tilde{s}(z) - \int_0^z dz \ \tilde{s}(z) .
  \label{sbraket}
\end{eqnarray}
Using these formulas, the stationarity equations  (see
eqs. (\ref{s-off}) and (\ref{s-on}))  become:
\begin{eqnarray}
\tilde{s}(z) = 2 \beta  f' \left({2 \over
  \beta}\left({1 \over  z( \mu+[\tilde{s}](z))}\ - \
 \int_z^1{dz \over z^2}\ {1 \over
\mu+[\tilde{s}](z)}\right)+{2 \over \beta}\sum_{\omega \neq 0} {1 \over m
\omega^2 +\mu  -\tilde{s}_d(\omega)} \right)
  \label{srsb}\\
\tilde{s}_d(\omega)=\int_0^1dz\ \tilde{s}(z)
- {2 \over \hbar} \int_0^U d\zeta \ (1-e^{i \omega\zeta})\
f'\left({2 \over \beta} \sum_{\omega' \neq 0} \ (1-e^{-i\omega'\zeta})
{1 \over m \omega'^2+\mu-\tilde{s}_d(\omega') }\right).
\label{sdrsb}
\end{eqnarray}

If we compare equation (\ref{srsb}) with the corresponding equation in
the classical case, we see that it is the same , apart from a
``renormalization'' or shift of the constant $a_0$ appearing in the
definition of the function $f$ (see eq. (\ref{f})):
\begin{eqnarray}
a_0 \rightarrow a_R(\beta,m,\mu,g)=a_0+b_0(\beta,\kappa,\mu,g).
  \label{ar}
\end{eqnarray}
The crucial point is that $b_0$ is independent of z.
We will discuss first the case of long-ranged correlations of the disorder.
One can repeat the steps carried out for the classical case
\cite{mp,engel}, and we find the following solution to eq.(\ref{srsb}):
\begin{eqnarray}
\tilde{s}(z)=\left\{ \begin{array}{ll}
 {3 \over 2}A z_1^2 & \hspace{0.5 in} 0<z<z_1 \nonumber \\
 {3 \over 2}A z^2  & \hspace{0.5 in}z_1<z<z_2 \nonumber \\
  {3 \over 2}A z_2^2  & \hspace{0.5 in}z_2<z<1 \end{array}\right.
  \label{rsb}
\end{eqnarray}
with
\begin{eqnarray}
A=(2/3)^3g^2\beta^3 \\
z_1={3 \over 2}g^{-2/3}\mu^{1/3}\beta^{-1}
  \label{rsbc}
\end{eqnarray}
and $z_2$ is the solution of the equation
\begin{eqnarray}
{1 \over2}\ \beta \ A \ a_R \ z_2^4\ +\ z_2\ -\ {3 \over 4}=0\ .
  \label{z2}
\end{eqnarray}

We observe that the only difference from the classical case is the
appearance of the renormalized function $a_R$ instead of $a_0$ in eq.
(\ref{z2}).
This solution becomes replica symmetric for $z_1=z_2$, which occurs
when
\begin{eqnarray}
  T_c=1/\beta_c={1 \over 2}g^{2/3}\mu^{-1/3}\ -\ {1 \over 2}
  a_R(\beta_c,m,\mu,g)
\label{Tcrsb}
\end{eqnarray}
This constitutes an equation for the transition temperature into the
glassy phase. Notice that for $\gamma=1/2$ this is the same as
equation (\ref{Tc}) in the previous section, which was obtained by
minimizing the mean-square displacement in the replica symmetric case.
Also, it follows from eqs. (\ref{x2d1}) and (\ref{rdz}) that in the
presence of RSB:
\begin{eqnarray}
\langle \langle{\bf x}^2 \rangle \rangle_R / N ={1 \over \beta \mu} +
{\tilde{s}(z=0) \over \beta \mu^2}+ {1 \over \beta \mu}\int_0^1\ {dz
\over z^2}{[\tilde{s}](z) \over \mu+[\tilde{s}](z)} + {1 \over \beta}
\sum_{\omega \neq 0} {1 \over m \omega^2+\mu-\tilde{s}_d(\omega)}.
\label{x2rsb}
\end{eqnarray}
Using the RSB solution obtained above, we find:
\begin{eqnarray}
\langle \langle{\bf x}^2 \rangle \rangle_R / N ={3 \over
2}g^{2/3}\mu^{-1/3} -{1 \over 2}a_R+{1 \over 2}b_0=
{3 \over 2}g^{2/3}\mu^{-1/3} -{1 \over 2}a_0,
\label{x2m2}
\end{eqnarray}
which is exactly the same as found for the minimum of the replica
symmetric expression in the previous section (eq.(\ref{x2m}) for
$\gamma=1/2$). Thus, through the entire glassy phase the mean-square
displacement remains locked at this constant value.

Let us comment briefly on the zero temperature limit of the RSB
solution. In that limit the solution given in eq. (\ref{rsb}) becomes
\begin{eqnarray}
{\tilde{s}(z) \over \beta}=\left\{ \begin{array}{ll}
 (g\mu)^{2/3} & \hspace{0.5 in} 0<z<z_1 \nonumber \\
 ({2 \over 3})^2(g\beta z)^2  & \hspace{0.5 in}z_1<z<z_2 \nonumber \\
  g a_R(\infty,m,\mu,g)^{-1/2}  & \hspace{0.5 in}z_2<z<1 \end{array}\right. ,
  \label{szt0rsb}
\end{eqnarray}
with
\begin{eqnarray}
z_1={3 \over 2}g^{-2/3}\mu^{1/3}\beta^{-1}\\
z_2={3 \over 2}g^{-1/2}a_R(\infty,m.\mu,g)^{-1/4}\beta^{-1}.
  \label{ut0}
\end{eqnarray}
For $\beta \rightarrow \infty$ this solution appears more
replica-symmetric like but it really isn't, since the value at z=0 is
always different from the value at $z \neq 0$ and the contribution
to the mean-square-displacement remains flat.

In the short-ranged case equation (\ref{srsb}) has a one-step RSB
solution, as in the classical case, again since the only difference
from the classical equation is a renormalization of $a_0$. The
equations can only be solved numerically \cite{engel}, and a detailed
investigation will be carried elsewhere.

\section{Concluding remarks}

In this paper we have investigated the combined effects of quantum
fluctuations and quenched disorder for the case of a particle subjected
to a combination of a fixed harmonic potential and a random potential
with power law spatial correlations. Acting separately, disorder and
quantum fluctuations both increase the mean-square-displacement of the
particle. From Figs. 2 and 4 it becomes evident that at high
temperature adding  quantum fluctuations in addition to disorder
increases the mean-square-displacement but at low temperature, in the
phase characterized by glassy behavior, the mean square displacement
remains locked probably due to tunneling effects among the different
free-energy minima.

We have seen that the phase diagram of the model is similar to that of
the quantum spin-glass in a transverse field. There is a transition at
zero temperature from a phase characterized by RSB effects to a phase
where such effects are not present.

It may be possible to extend the problem of a quantum
particle, to higher-dimensional manifolds in a disordered medium, and
explore the effects of quantum fluctuations. It may be also possible
to extend the problem to many particle systems where statistics might
play an  important role. We hope our work will stimulate
further research in this interesting area.

\acknowledgements
I thank Dr. H. A. Duncan for useful discussions and Mr. Hsuan-Yi Chen
for  checking some of the algebra and performing the numerical
calculation reported in Appendix C.

\newpage
\appendix
\section{}

The major step in calculating the variational free-energy is
evaluating the expectation value:
\begin{eqnarray}
  \left\langle f\left({({\bf x}_a(u)-{\bf x}_b(u'))^2\over
    N}\right)\right\rangle_{h_n}.
\label{fexp}
\end{eqnarray}
This is done by expanding $f$ in a power series and using the
following formula:
\begin{eqnarray}
\left\langle( ({\bf x}_a(u)-{\bf
  x}_b(u'))^{2j}\right\rangle_{h_n}={(N+2j-2)!! \over (N-2)!!}
(\hat{G}_{aa}(0) +\hat{G}_{bb}(0)-2\hat{G}_{ab}(u-u'))^j\ ,
  \label{powerexp}
\end{eqnarray}
where we defined
\begin{eqnarray}
N \hat{G}_{ab}(u-u')=\left\langle{\bf x}_a(u)\cdot{\bf
  x}_b(u')\right\rangle_{h_n}.
  \label{Gdef}
\end{eqnarray}
Resumming the series gives rise to
$\hat{f}(\hat{G}_{aa}(0)+\hat{G}_{bb}(0)-2\hat{G}_{ab}(u-u'))$ .

In momentum space conjugate to the Trotter time the variational
free-energy becomes:
\begin{eqnarray}
n \beta \left< F \right>_R /N \ = \ {1 \over 2} \sum_\omega \sum_{ab}
G_{ab}(\omega) \tilde{s}_{ab}(\omega) -{1 \over 2} \sum_\omega
{\rm tr}\ \ln G(\omega) \nonumber \\
 + {\beta \over 2 \hbar} \sum_{ab} \int_0^U d \zeta \
 \hat{f} \left({1 \over \beta} \sum_{\omega'}\
(G_{aa}(\omega')+ G_{bb}(\omega')-
2 e^{-i\omega'\zeta} \ G_{ab}(\omega'))\right) + const.
  \label{varf}
\end{eqnarray}
where $\tilde{s}_{ab}(\omega)/\beta$ is the Fourier transform of
$s_{ab}(u-u')$ and $G_{ab}(\omega)$ os the Fourier transform of
$\hat{G}_{ab}(u-u')$.
This free-energy coincides with the one derived from the large-$N$
limit with the replacement $f\rightarrow \hat{f}$.

\section{}

In this Appendix we provide some further details on the numerical
solution of equations (\ref{s-rs}) and the evaluation of the
mean-square-displacement, eq. (\ref{x2f1}).
Our goal is to minimize the error in the truncation of high
frequencies. Hence we define the function $t_d(\omega)$
for $\omega \neq 0 $by
\begin{eqnarray}
\tilde{t}_d(\omega_l)=\tilde{s}_d(\omega)+\tilde{t}
  \label{td}
\end{eqnarray}
Where $\tilde{t}$ will be adjusted such that for the highest frequency
included, $\tilde{t}_d(\omega_{max})=0$.
We then use the formula \cite{gr}:
\begin{eqnarray}
{1 \over \beta}\sum_{\omega \neq 0}{e^{-i \omega \zeta}  \over m
  \omega^2+\mu} =
{1 \over 2 \sqrt{\kappa \mu}} {\cosh(\alpha(1-2\zeta/U)) \over
  \sinh(\alpha)}- {1 \over \beta \mu}
  \label{sumfor}
\end{eqnarray}
with
\begin{eqnarray}
\alpha={\beta \over 2} \sqrt{{\mu \over \kappa}},
  \label{alpha}
\end{eqnarray}
to write
\begin{eqnarray}
{2 \over \beta}\sum_{\omega \neq 0}{e^{-i\omega\zeta} \over
  m\omega^2+\mu-\tilde{s}_d(\omega)}= - {2 \over \beta \mu_R}+{1 \over
   \sqrt{\kappa \mu_R}} {\cosh(\alpha_R(1-2\zeta/U)) \over
  \sinh(\alpha_R)} \nonumber \\ +{4 \over \beta}
\sum_{k=1}^{k_m}{e^{-i\omega_k\zeta}\tilde{t}_d(\omega_k) \over
  (m\omega_k^2+\mu+\tilde{t})(m\omega_k^2+\mu+\tilde{t}-\tilde{t}_d(\omega_k)})
  \label{arg}
\end{eqnarray}
where
\begin{eqnarray}
\mu_R=\mu+\tilde{t}\\
\alpha_R={\beta \over 2} \sqrt{{\mu_R \over \kappa}}.
  \label{mur}
\end{eqnarray}
A similar expression is used for the case of $\zeta=0$.
The equations for $\tilde{t}_d(\omega_k)$, $\tilde{t}$ and $\tilde{s}$
were solved numerically with $k_{max}=10$. We used a fortran routine
\cite{recipes}
that finds a root for a set of non-linear equations. The solution was
then used to calculate the mean-square displacement.

\section{}

In this Appendix we give some of the details omitted in
ref. \cite{prl} for lack of
space, as well implement some corrections. Some of the notation is
slightly different than in the present
paper. Starting with eq. (\ref{F&J}) we proceeded in (I) to look for a
saddle point under the
constraint that the off-diagonal elements of the order parameters $r$
and $s$ are independent of the Trotter time, whereas the diagonal
elements denoted by $\chi(u-u') \equiv r_d(u-u')$ and $\nu(u-u')
\equiv s_d(u-u')$ are dependent on $u-u'$.
In the 1-step RSB we have used the notation
\begin{eqnarray}
r(z)=r_2, \ \ \ z<k \\
r(z)=r_{11}, \ \ z>k
\end{eqnarray}
where k has been used to denote the breaking point $z_c$. A similar
notation apply for $s$. Substituting these order parameters in the
expression for the free-energy we obtain:
\begin{eqnarray}
\beta \left< F \right>_R / N\ = \ {\beta^2 \over 2}[k(r_2 s_2- r_{11}
s_{11})-r_2 s_2]\ +\ {1 \over 2 \hbar^2} \int_0^U
\int_0^U du du' \chi(u-u') \nu(u-u')\nonumber \\
-{\beta^2 \over 2 }k f(2\chi(0)-2r_{11})\ + \
{\beta^2 \over 2 \hbar^2} (k-1) f(2\chi(0)-2r_2) \nonumber \\
+{1 \over 2 \hbar^2}\int \int du du'\ f(2\chi(0)-2\chi(u-u'))-{1
  \over k} \int_{-\infty}^\infty Dz \ \ln \int Dy \ (Z_H)^k + const.
\end{eqnarray}
where $Dz \equiv (dz/\sqrt{2\pi})\exp(-z^2/2)$ and similarly $Dy$, and
$Z_H \equiv \int dx \int [dx]e^{-H/\hbar}$. H is the one-dimensional,
one-particle effective hamiltonian given by:
\begin{eqnarray}
H=\int_0^U du \left [{m \over 2} \dot{x}^2(u) + {\mu
\over 2} x^2(u) -(z\sqrt{s_{11}}+y\sqrt{s_2-s_{11}})x(u)\right]
\nonumber \\
-{1 \over 2\hbar}\int_0^U \int_0^U du du'\left(\nu(u-u')-s_2\right)
x(u)x(u')
\label{H}
\end{eqnarray}

The different order-parameters are determined self-consistently from
the equations which extremize $\left<F\right>_R$:
\begin{eqnarray}
\chi(u-u')\ = \ \int Dz \ {\int Dy \left<x(u)x(u')\right>_H Z_H^k \over
\int Dy Z_H^k}, \label{chi} \\
r_{11}\ =\ \int Dz \ \left({\int Dy \left<x(u)\right>_H Z_H^k \over
\int Dy Z_H^k}\right)^2, \\
r_2 \ = \ \int Dz \ {\int Dy \left<x(u)\right>_H^2 Z_H^k \over
\int Dy Z_H^k}, \\
\nu(u-u') = [2 k f'(2\chi(0)-2r_{11})+2 (1-k)
f'(2\chi(0)-2r_2)\nonumber\\
-{2 \over U^2} \int_0^U\int_0^U du du' f'(2\chi(0)-2\chi(u-u'))]U
  \delta (u-u') +2 f'(2\chi(0)-2\chi(u-u'))\\
s_{11}\ = \  2 f'(2\chi(0)-2r_{11}), \\
s_2\ = \  2 f'(2\chi(0)-2r_2).
\end{eqnarray}
The equation for the breaking point $k$ is given by
\begin{eqnarray}
 \ {\beta^2 \over 2}(r_2 s_2- r_{11}
s_{11})
-{\beta^2 \over 2} f(2\chi(0)-2r_{11})\
+{\beta^2 \over 2} f(2\chi(0)-2r_2) \nonumber \\
+{1 \over k^2} \int_{-\infty}^\infty Dz \ \ln \int Dy \ (Z_H)^k
- {1 \over k} \int Dz  {\int Dy (Z_H)^k \ln Z_H \over \int Dy (Z_H)^k}=0.
\end{eqnarray}

We have continued by putting the ``time'' variable on a lattice, with lattice
spacing $U/M$, where $M$ is to be taken eventually to infinity. In
practice we carried out exact calculations up to values of $M=20$, and
extrapolation to $M=\infty$ has been made by finite size scaling and the
$1/M^2$ rule \cite{yglai}. The calculation involves finding the inverse and
the determinant of the $M \times M$ matrix ${\cal M}$ appearing
in the expression for the discretized effective hamiltonian, see eq. (\ref{H}):
\begin{eqnarray}
{\cal M}_{i,j}\ =\ \delta_{ij} (2\kappa{M \over
\beta}+{\mu\beta \over M})-{\kappa M \over \beta}\delta_{i+1,j}-
{\beta^2 \over M^2}(\nu(|i-j|)-s_2) \hspace{.2in} i\leq j, i\neq 1,
\label{calM}
\end{eqnarray}
where $\kappa \equiv m/\hbar^2$. For $i=1$ there is an additional term
$(-\kappa M / \beta)\delta_{M,j}$. For $j <  i$ the matrix is given
by symmetry.

Using this notation, the final result for the discretized free-energy is:
\begin{eqnarray}
\beta \left< F \right>_R / N\ = \ {\beta^2\over 2} [k(r_2 s_2- r_{11}
s_{11})-r_2 s_2]\ +\ {\beta^2 \over 2M^2} \sum_{ij} \chi(|i-j|)
\nu(|i-j|)\nonumber \\
-{k\beta^2 \over 2 }  f(2\chi(0)-2r_{11})\ + \
{(k-1)\beta^2 \over 2} f(2\chi(0)-2r_2)+{\beta^2 \over 2 M^2}
\sum_{ij} f(2\chi(0)-2\chi(|i-j|)) \nonumber \\
-{M \over 2}\ln \left({\kappa M \over \beta}\right) +{1 \over 2} \ln\
\det \ {\cal M} +{1 \over 2k} \ln (1-k\sigma v_2^2)- {\sigma v_1^2 \over
2(1-k\sigma v_2^2)}\ ,
\label{discF}
\end{eqnarray}
where we defined
\begin{eqnarray}
\sigma={1 \over M} \sum_j ({\cal M}^{-1})_{ij},\hspace{0.1in}
v_1^2=\beta^2 s_{11}, \hspace{0.1in} v_2^2=\beta^2 (s_2-s_{11}).
\label{sig}
\end{eqnarray}
The different order parameters have been obtained by extremizing
eq. (\ref{discF}).

\newpage

{\bf Figure captions:}

Fig. 1: Schematic phase diagram of a quantum particle in a random
potential plus a fixed harmonic potential.

Fig. 2: Plot of the mean-square-displacement vs. $t=T/T_c(\infty)$ for
the long-ranged case ($\gamma=1/2$). Curves from right to left are for
$\kappa= 1000,2,1,0.5,0.25,0.2,0.18,0.1$. The horizontal dashed line represent
the RSB
solution below $T_c(\kappa)$.

Fig. 3: Phase diagram for the long-ranged case for $g=2\sqrt{2}$ and
$\mu=1$. The dashed line is a possible extrapolation to $T=0$.

Fig. 4: Plot of the mean-square-displacement vs. $T$ for
the short-ranged case ($\gamma=3/2$). Notation as in Fig. 2.

Fig. 5: Phase diagram for the short-ranged case for $g=1/\sqrt{2\pi}$
  and $\mu=0.53$. Curves from right to left are for
  $\kappa=1000,10,4,3,2.5,2,1$.

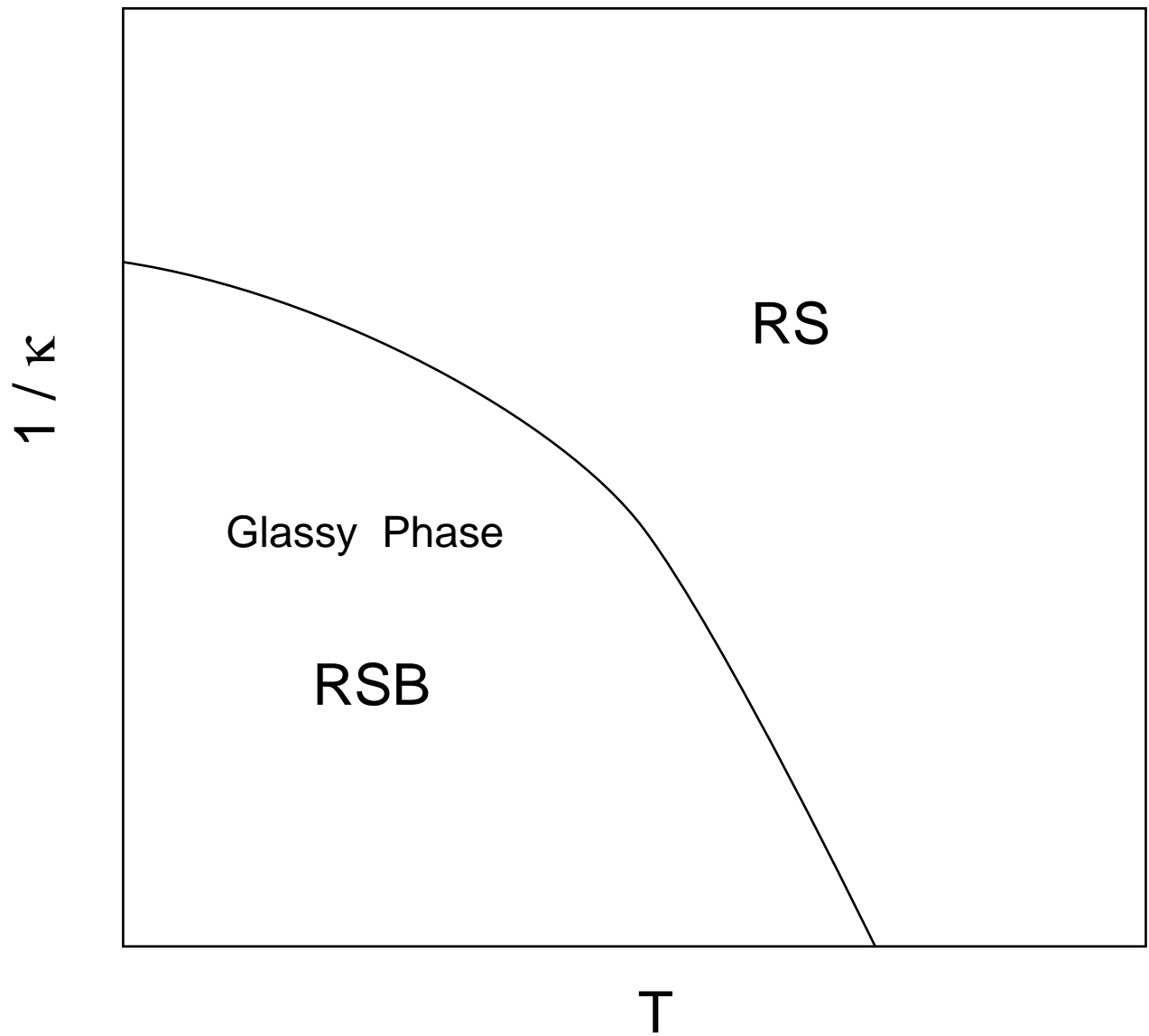
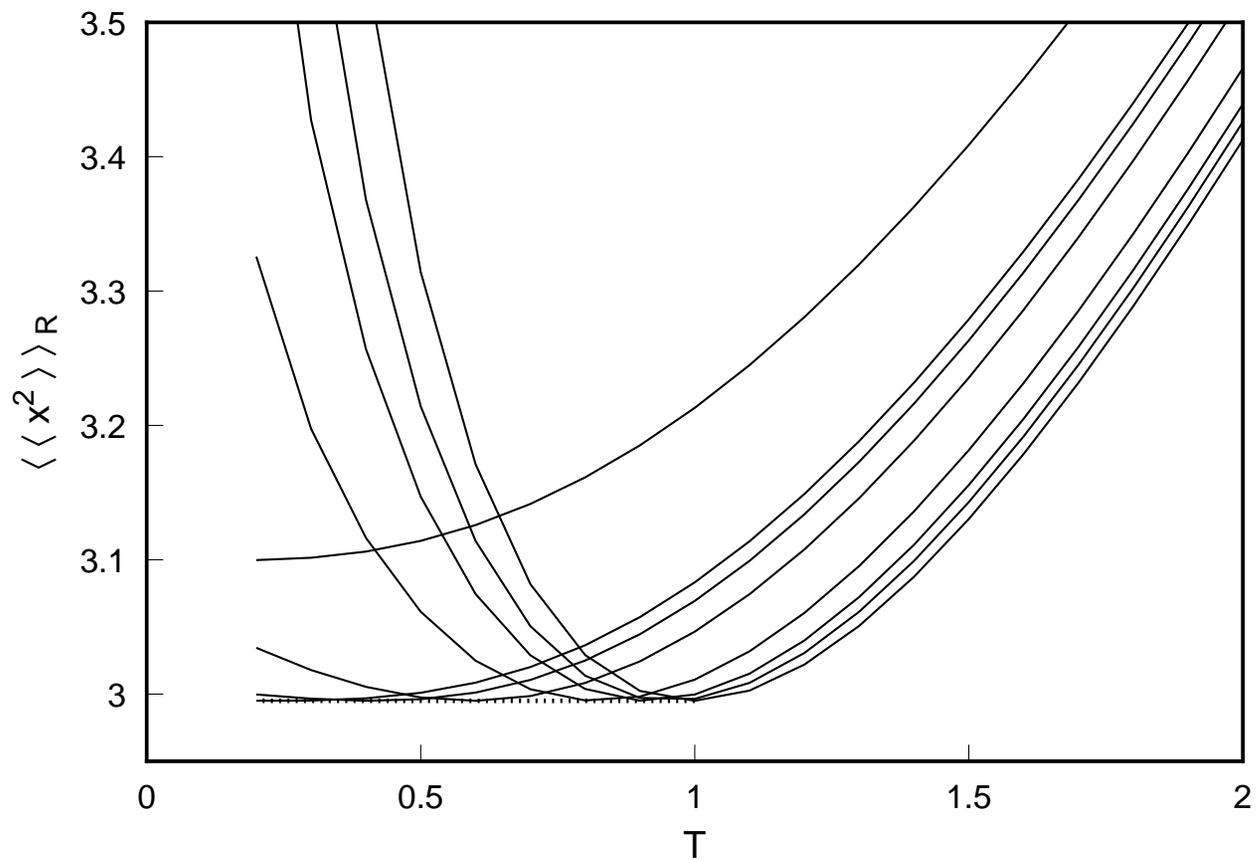
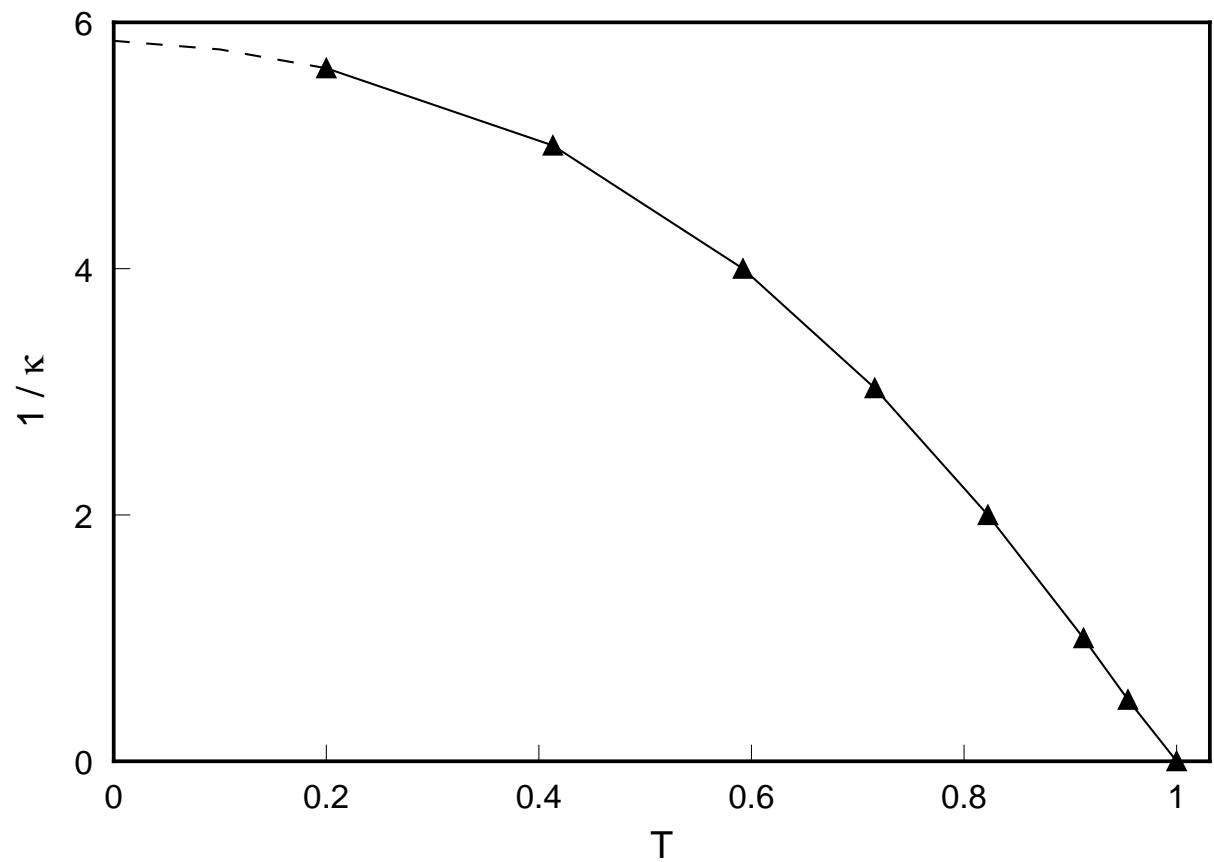
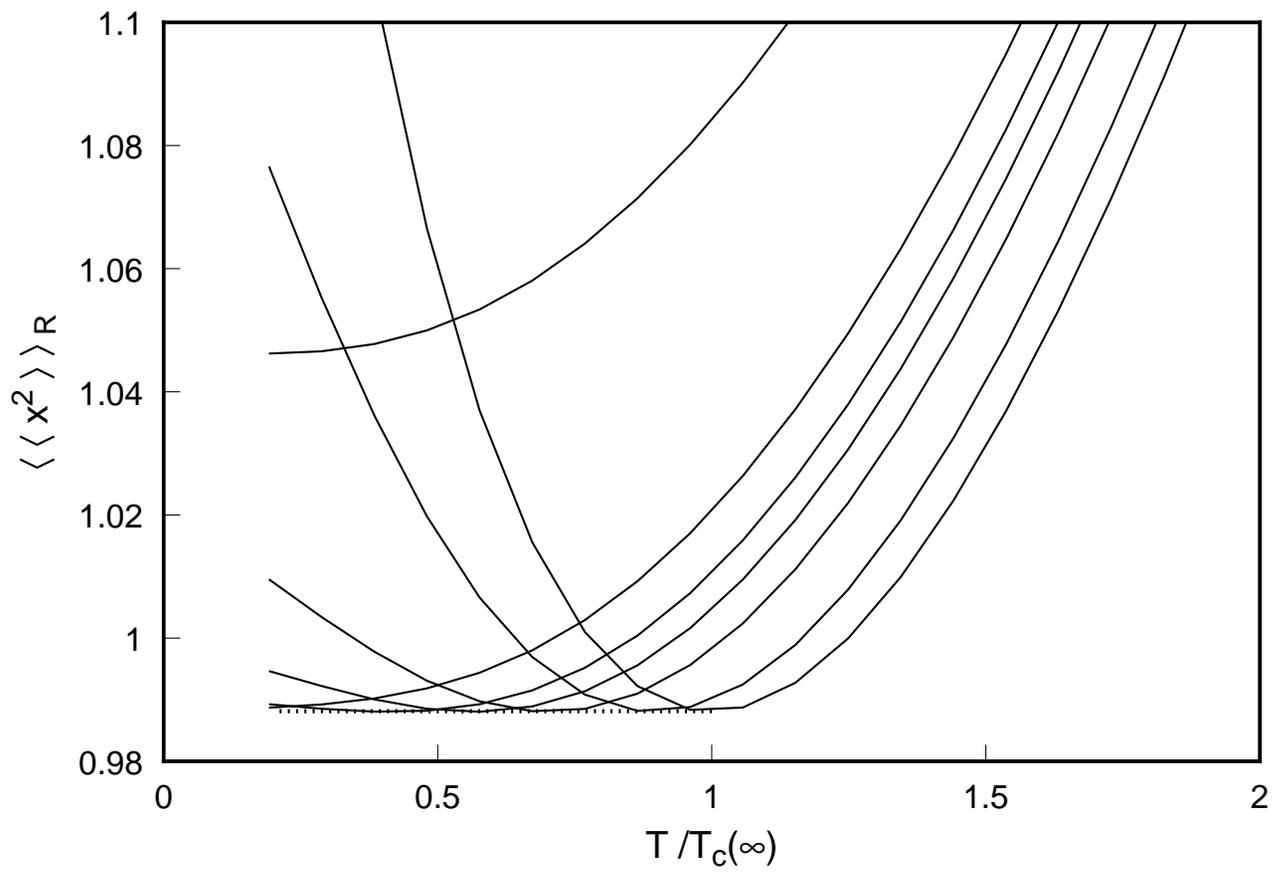
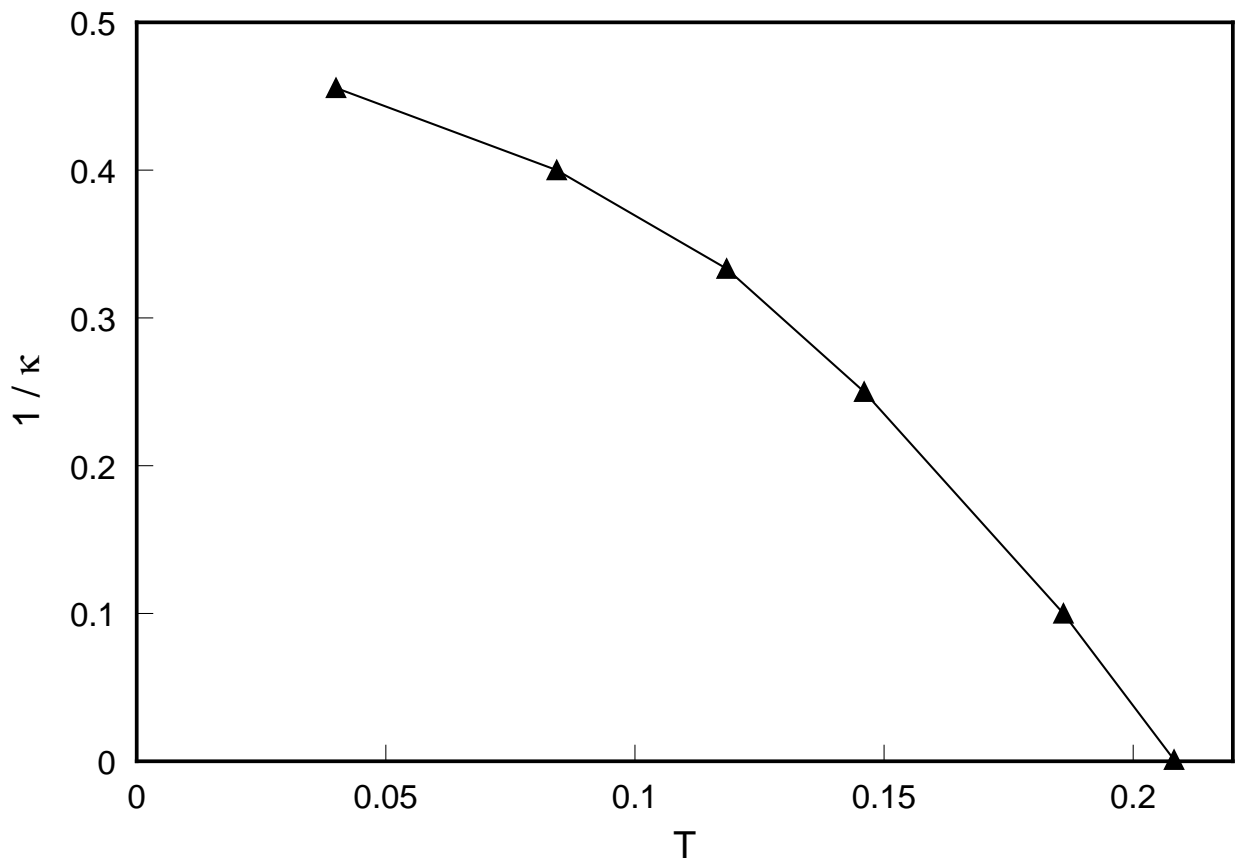
\end{document}